\documentclass[12pt]{iopart}
\usepackage{graphicx}% Include figure files
\usepackage{subfigure}
\usepackage{epsfig}
\usepackage{cite}
\begin{document}

\title[]{Shape evolution of MBE grown  Si$_{1-x}$Ge$_{x}$ structures on high index Si(5 5 12) surfaces: A temperature dependent study}
\author{J. K. Dash, A. Rath, R. R. Juluri and P. V. Satyam}
\address{Institute of Phsyics,  Sachivalaya
Marg, Bhubaneswar - 751005, India}
\eads{\mailto{pvsatyam22@gmail.com}, \mailto{satyam@iopb.res.in}}

\begin{abstract}
The morphological evolution and the effect of growth temperature
on size, orientation and composition of molecular beam epitaxy
grown Ge-Si islands on Si(5 5 12) surfaces have been investigated in the  temperature range from room
temperature (RT) to 800$^\circ$C. Two modes of substrate
heating i.e. radiative heating (RH) and direct current heating
(DH) have been used. The post-growth characterization was carried out
ex situ by scanning electron microscopy (SEM), cross-sectional transmission electron microscopy (X-TEM) and
Rutherford backscattering spectrometry (RBS). In the RH case, we
found spherical island structures  at 600$^\circ$C with a bimodal
distribution and upon increasing temperature, the structures got faceted at
700$^\circ$C. At 800$^\circ$C  thick ($\sim$ 122nm) dome like
structures are formed bounded by facets. While in the case of DC heating, after the optimum
critical temperature 600$^\circ$C, well aligned trapezoidal
Si$_{1-x}$Ge$_x$ structures with a graded composition starts
forming along the step edges. Interestingly, these aligned structures have been found only around 600$^\circ$C, neither at low temperature nor at higher temperatures.
\end{abstract}


\maketitle

\section{Introduction}

Self-assembly of strain-induced islands in hetero-epitaxial systems
is a promising route  to use nanoscale structures as quantum dots
in opto-electronic devices. Among the different strained material
combinations investigated so far, the Ge/Si system is often
considered as a prototype for understanding fundamental properties
of heteroepitaxial growth. In general, Ge prefers to adsorb on the
surface, since the Ge dangling-bond energy is lower than that of
Si \cite{nakajima,uberuaga}. However, theoretically, it is
expected that the tensile stress induced due to reconstruction
makes Ge atoms diffuse into the silicon subsurface and into the
bulk \cite{cho}. Hence, evolution of crystallographic facets of
strained Ge-Si structures and diffusion of Ge into the Si
substrate depend on the kind of the substrate and nature of
reconstructions\cite{robinson}.

Faceting is a fundamental concern in the crystal growth process. It is observed when the system is
allowed to minimize its free energy (thermodynamic
equilibrium)\cite{bermond,herring,saul}.The faceting phenomenon not
only  depends on material properties but also on the sample
configurations and process conditions (pressure, growth
temperature, composition, etc.). Faceting can  be used in order to
improve electrical performance as in folded devices  or minimized
and suppressed as in silicon on nothing (SON) technology
\cite{pribat}.

The growth of Ge nanostructures on low index silicon surfaces has been studied by many groups\cite{legoues94,mo90,bert01,pcti10} in the last two decades. More recently, the study of self-organization in Ge/Si system has been extended to high-index surfaces\cite{kim08,kim12,omi99,dash11}. However, the kinetic and
thermodynamic factors determining the shape of the self-assembled Ge-Si structures  are not yet well studied.

A high index surface is misoriented from the low index plane by a small vicinal angle, relative to low index surfaces. The ultra-clean reconstructed surface shows periodic steps and terraces. The anisotropic surface reconstruction on high-index silicon surfaces makes them potentially significant substrates for the growth of ordered nanostructures. The use of such matrices as templates can enhance the formation of aligned nanostructures with self-assembled growth, which is not so easily achievable with lithography tools. Due to its natural process and damage-free features, structural self-organization on high index surfaces has received much attention in the formation of coherent 1D and 2D semiconductor / metal nanostructures. High index surfaces are found to have reconstructions into regular hill (step) and valley (terrace) structures with periods ranging from several nanometers to about one hundred nanometers. The periodicity of such structures depends mainly on the substrate orientation and miscut or vicinal angles. The width of the terraces and step height can easily be modulated with temperature and coverage. These anisotropic silicon surfaces having alternating terraces and atomic steps can be used to form aligned one dimensional nanostructures\cite{bas95,bas01,ahn02}. Among the high index silicon surfaces oriented between (001) and (111), Si(5 5 12) exhibits relatively stable reconstruction having 1D symmetry with a ($1\bar{1}0$) mirror plane. The Si(5 5 12) surface is oriented 30.5$^\circ$ away from (001) towards (111) and offers a stable 2$\times$1 surface reconstruction with one-dimensional periodicity over a large unit cell\cite{bas95,kim07}.

In this paper, we investigate the morphological
evolution of strained Ge-Si islands on high index Si(5 5 12)
surfaces  as a function of growth temperature and mode of heating
the substrate. We show the transformation of irregular shaped
Ge-Si island to faceted dome structures and well aligned
trapezoidal structure after an optimum critical temperature
depending upon the modes of heating the substrate.

\section{Experimental details}
The experiments discussed in the following were performed in an
MBE chamber under ultra-high vacuum(UHV), at a base pressure of
$\sim 2.5 \times 10^{-10}$ mbar. Si(5 5 12) of size 8$\times$3 (mm)$^2$ were prepared through
cutting from commercially available p-type boron doped wafers (of
resistivity of 10 - 15  $\Omega$ cm). Substrates were degassed at
600$^{\circ}$C for about 12 hours followed by repeated flashing
(with direct current heating) for 30 sec. at a temperature of
1250$^{\circ}$C to remove the native oxide layer to obtain a clean
and well-reconstructed surface. The reconstruction has been
confirmed with in-situ reflection high energy electron diffraction
(RHEED). The temperature was monitored with an infra-red pyrometer
calibrated with a thermocouple attached to the sample holder. Two
series of samples have been prepared. 10 ML Ge deposition on Si(5
5 12) from RT to 800$^{\circ}$C, with two modes of heating,
namely, (i) heating is achieved through a filament underneath
(radiative heating: RH). (ii) heating achieved by passing a direct
current (DC) through the sample. The post-growth characterization
of the samples was carried out ex situ by field emission gun based
scanning electron microscopy (SEM), transmission electron
microscope(TEM) and MeV ion scattering (Rutherford backscattering
spectrometry (RBS)). To compare with growth on a high index surface, we also carried out 10 ML thick
Ge growth on a Si(111) substrate with a resistivity of 0.5 - 30 $\Omega$ cm. The base pressure inside the MBE chamber was
$\sim 3.5 \times 10^{-10}$ mbar. A direct current of 1.5 A (5.7 V) as applied along the $\langle 1 1 \bar{2}\rangle$ direction. The post-growth
characterization of the samples was done ex situ by field emission gun based SEM.

\section{Results and discussions}
In this report, we will emphasize on the shape transformations of
Ge-Si structures on clean Si(5 5 12) surfaces as a function of
substrate temperature in two modes of heating condition, i.e by radiative heating (RH) and direct current heating.

\subsection{Shape evolution in RH condition}

\begin{figure}
\centering \vspace{0.4cm} \epsfig{file=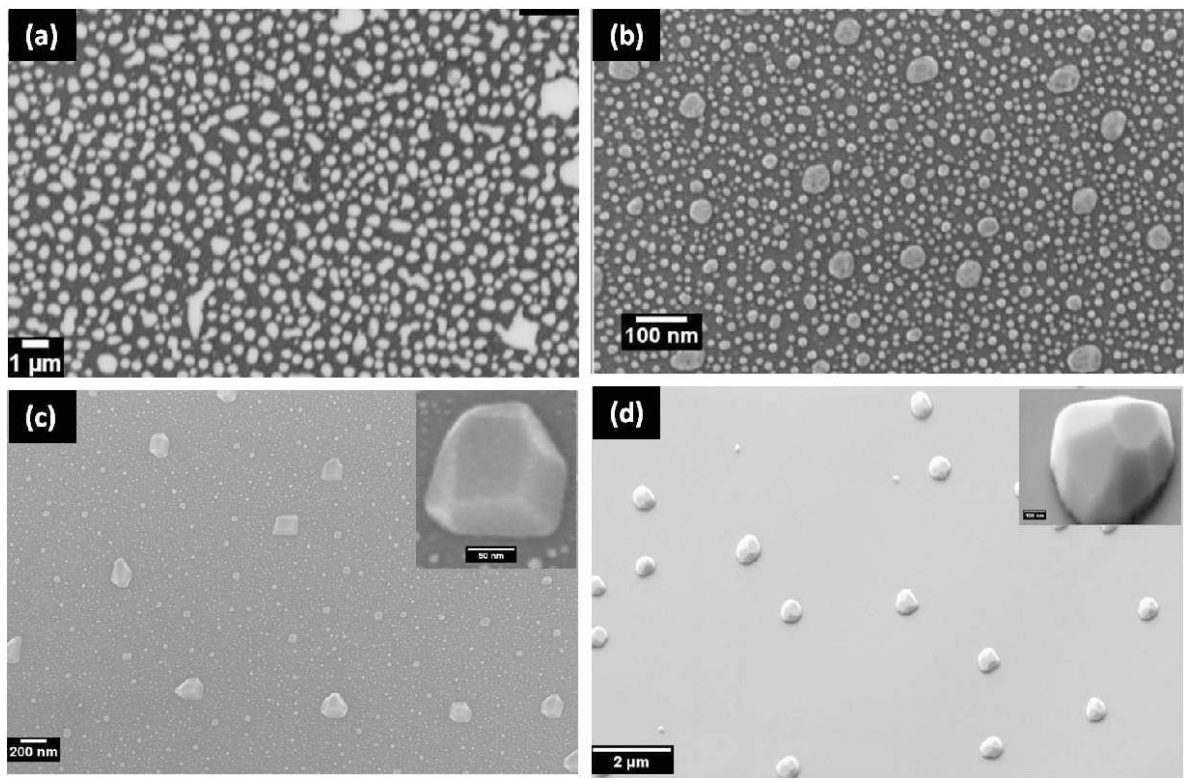, width=\linewidth}
\caption{SEM micrographs  of 10 ML Ge/Si(5 5 12) at (a) RT
(b) 600$^{\circ}$C - RH (c) 700$^{\circ}$C - RH
(d) 800$^{\circ}$C - RH.}\label{fig1}
\end{figure}

\begin{figure}
\centering \vspace{0.4cm} \epsfig{file=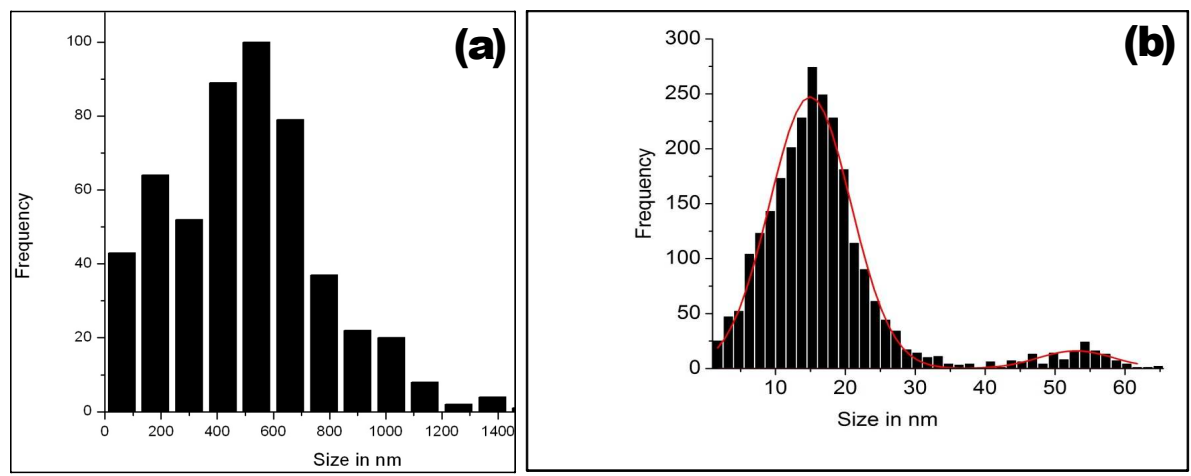, width=\linewidth}
\caption{Histogram of (a) island size in 10 ML ge/Si(5 5 12) system at RT and (b) bimodal size distribution in 10 ML ge/Si(5 5 12) in RH condition.}\label{fig2}
\end{figure}

\begin{figure}
\centering  \vspace{0.4cm} \epsfig{file=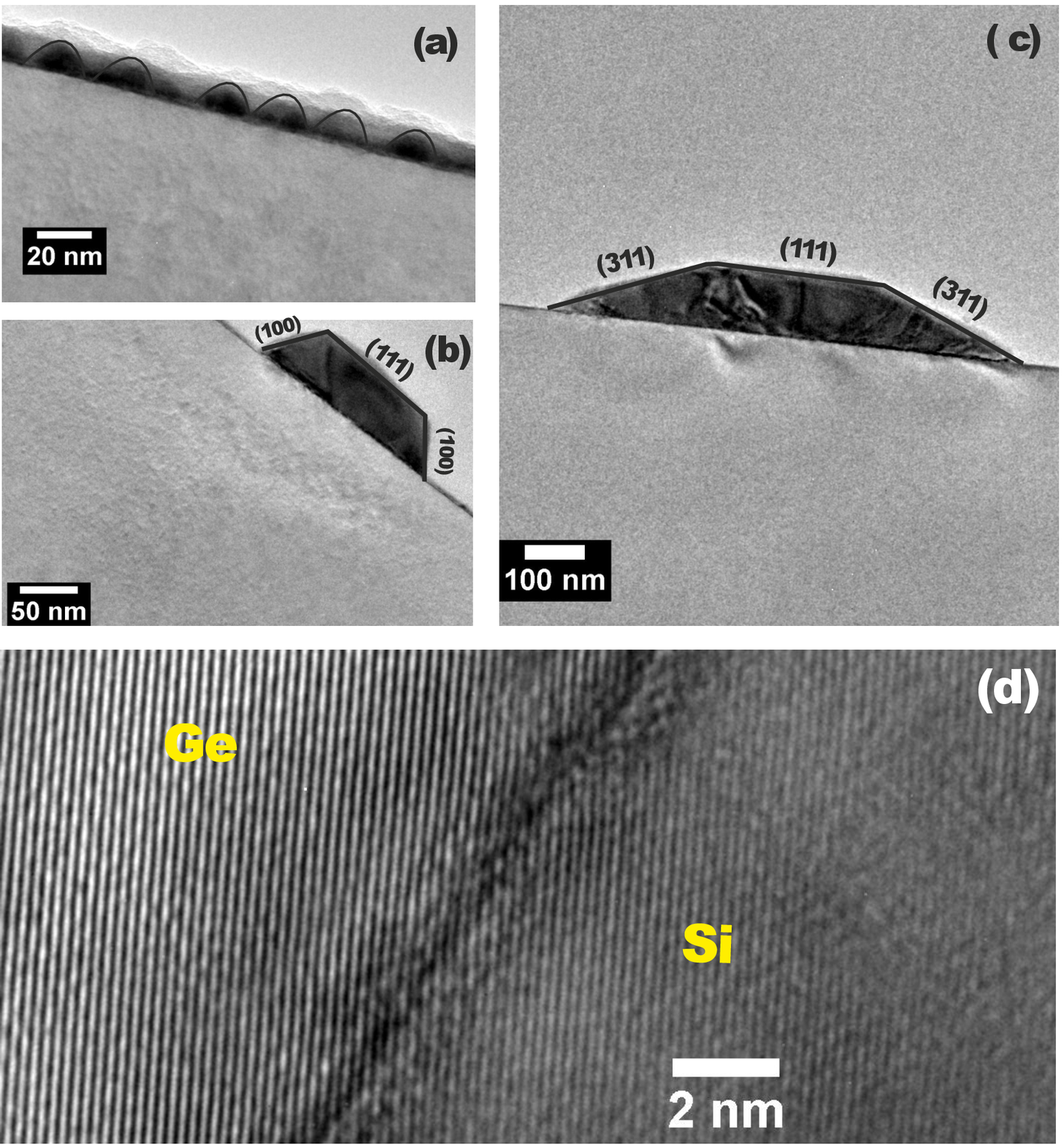, width=\linewidth}
\caption{Cross-sectional TEM images of 10 ML Ge/Si(5 5 12) at (a)
 600$^\circ$C - RH (the marked line is to show a single island) (b)faceted island at 700$^\circ$C - RH (c) faceted dome structures at
 800$^\circ$C - RH. (d) showing the epitaxial nature of faceted Ge-Si structures.}\label{fig3}
\end{figure}

 Fig. \ref{fig1} shows SEM micrographs for 10 ML Ge growth on Si(5 5
12) surface at various substrate temperatures in RH condition.
Fig. \ref{fig1} (a) shows the FEGSEM image for  Ge film deposited, while keeping the substrate at room temperature (RT). The island
size distribution of the random structures are given in fig. \ref{fig2} (a). The mean size of the islands is found to be 504 $\pm$ 38 nm in
diameter. Fig. \ref{fig1} (b) depicts the morphology for
Ge film deposited at a substrate temperature $T= 600^\circ$C in RH condition. In
this case, a distribution of bigger spherical Ge islands
surrounded by a large number of smaller islands was observed,
showing a kind of Ostwald ripening process with a bimodal
distribution of islands [fig. \ref{fig2} (b)]. The mean value for the size of the
smaller islands was  found to be 15.8 $\pm$ 1.5 nm, while the
bigger islands are found to have an average size of 54.0 $\pm$ 7.0
nm. For the 700$^\circ$C deposition case, the mean size of the
island increases at the expense of decreasing the number of
smaller islands. I. Daruka \emph{et al} \cite{daruka99} showed the
shape evolution of islands to various faceted structures in terms
of surface energy minimization with increasing volume. Normally, the
chemical potential of an island decreases continuously with size,
due to the smaller surface to volume ratio. As a result, material
diffuses from smaller to larger islands. In this coarsening or
"Ostwald ripening", large islands grow while small islands shrink
and disappear.  Above a lower transition volume, the Ge-Si island
is stable and got faceted, as seen in fig. \ref{fig1}(c). The mean size of the faceted
island is 150 $\pm$ 15 nm. To get a clearer view of this facet
structures, cross-sectional TEM (X-TEM) analysis has been carried out. Fig. \ref{fig3} depicts the interfacial morphology of these Ge structures grown on Si(5 5 12) under RH conditions. In fig. \ref{fig3}(a), the X-TEM image of spherical islands formed at $T= 700^\circ$C - RH has been displayed. Fig. \ref{fig3}(b) shows a cross-sectional view of faceted island
structure formed in the $T = 700^\circ$C case, where the thickness
of the island is 42.6 $\pm$ 2.3 nm. A set of Ge/Si(5 5 12) sample
has been prepared keeping $T = 800^\circ$C. Upon increasing the
temperature, it causes the formation of only larger dome like
structure with disappearance of smaller islands. Fig. \ref{fig1}(d) displays the Ge-Si
dome like structures bounded by facets with average size of 630.5 $\pm$ 16.4 nm. The thickness
of the faceted dome structure is found from the cross-sectional
TEM view in fig. \ref{fig3}(c), which is 122 $\pm$ 5.6 nm. We have also calculated the angles between the faceted planes and determined the orientation of facet planes as per the study done by L.D. mark\cite{ldmark94}. In the case of $T = 700^\circ$C - RH, the faceted island is bounded by $\left\{111\right\}$ and $\left\{100\right\}$ planes and the faceted dome structure in the $T = 800^\circ$C - RH case, are bounded by $\left\{111\right\}$ and $\left\{311\right\}$ set of planes. Fig. \ref{fig3}(d) confirms that the faceted Ge-Si structures are epitaxial in nature.

\begin{figure}
\centering \vspace{0.4cm} \epsfig{file=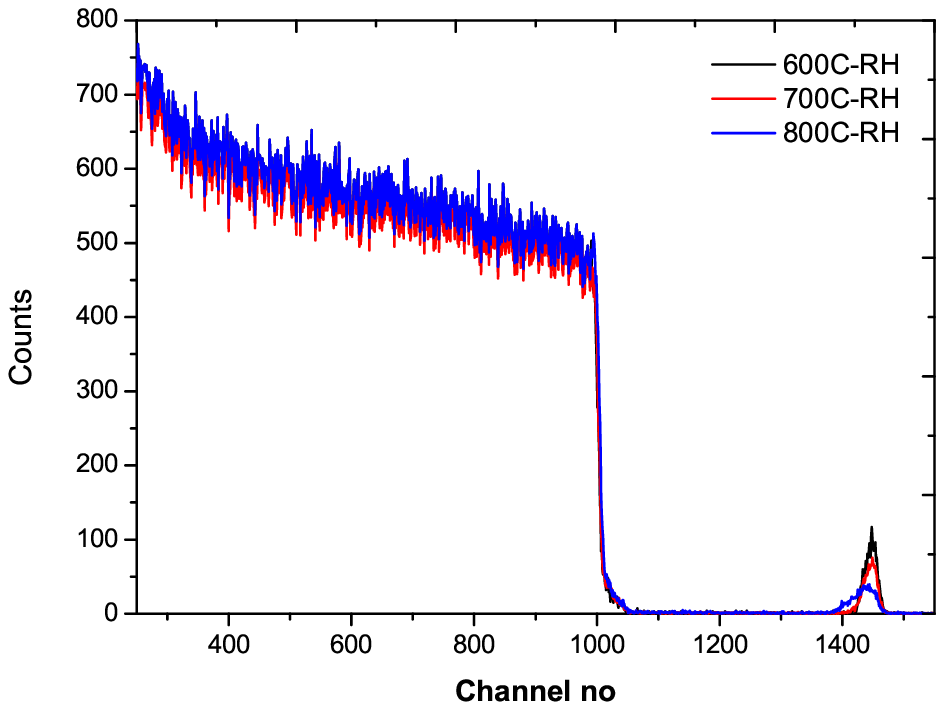, width=\linewidth}
\caption{Experimental RBS spectrum from 10 ML Ge/Si (5512) at T =
600$^\circ$C, 700$^\circ$C and 800$^\circ$C in RH
case.}\label{fig4}
\end{figure}

S.A. Chaparro et al \cite{chaparro} showed that higher growth
temperature activates additional pathways for the Ge islands to
relieve their strain via Ge/Si intermixing. Si-Ge alloying causes
the formation of quite large hut clusters for $T>600^\circ$C.
Furthermore, they observe that there is an increase in the mean
dome cluster size with increasing temperature. RBS experiment has been performed on all the three sets of sample i.e.
Ge/Si(5 5 12) at $T$ = 600$^\circ$ C, 700$^\circ $C and 800$^\circ
$C to investigate the possible intermixing of Si-Ge. It is evident
from fig. \ref{fig4} that there is Ge diffusion towards silicon
for $T $= 700$^\circ$ C and $T $= 800$^\circ$C. The simulated composition of Ge
and Si are tabulated in table \ref{tab:table1}. Ge-Si alloying may
also be attributed in the strain relief mechanism in forming the
larger facets structures minimizing the surface energy.

\begin{table}
\caption{\label{tab:table1} Parameters obtained on fitting the
experimental RBS data for Ge/Si(5 5 12) at T=800$^\circ$C - RH,
using the SIMNRA simulation code.}
\begin{indented}
\item[]\begin{tabular}{@{}llll}
\br
 Layer &Ge  &Si  &Thickness \\
number&composition&composition&(1$\times$10$^{15}$
atoms.cm$^{-2}$)\\
\mr
 1 & 0.6000 & 0.4000 & 0.100\\
 2& 0.0600 & 0.9400 & 100.0\\
 3& 0.0075 & 0.9925 & 250.0\\
 4&0.0070 & 0.9930 & 300.0\\
 5& 0.0050 &0.9950 & 350.0\\
 6& 0.0020 & 0.9980 & 400.0\\
 7& 0 & 1 & Bulk\\
 \br
\end{tabular}
\end{indented}
\end{table}

\subsection{Shape evolution in DH condition}

\begin{figure}
\centering  \vspace{0.4cm} \epsfig{file=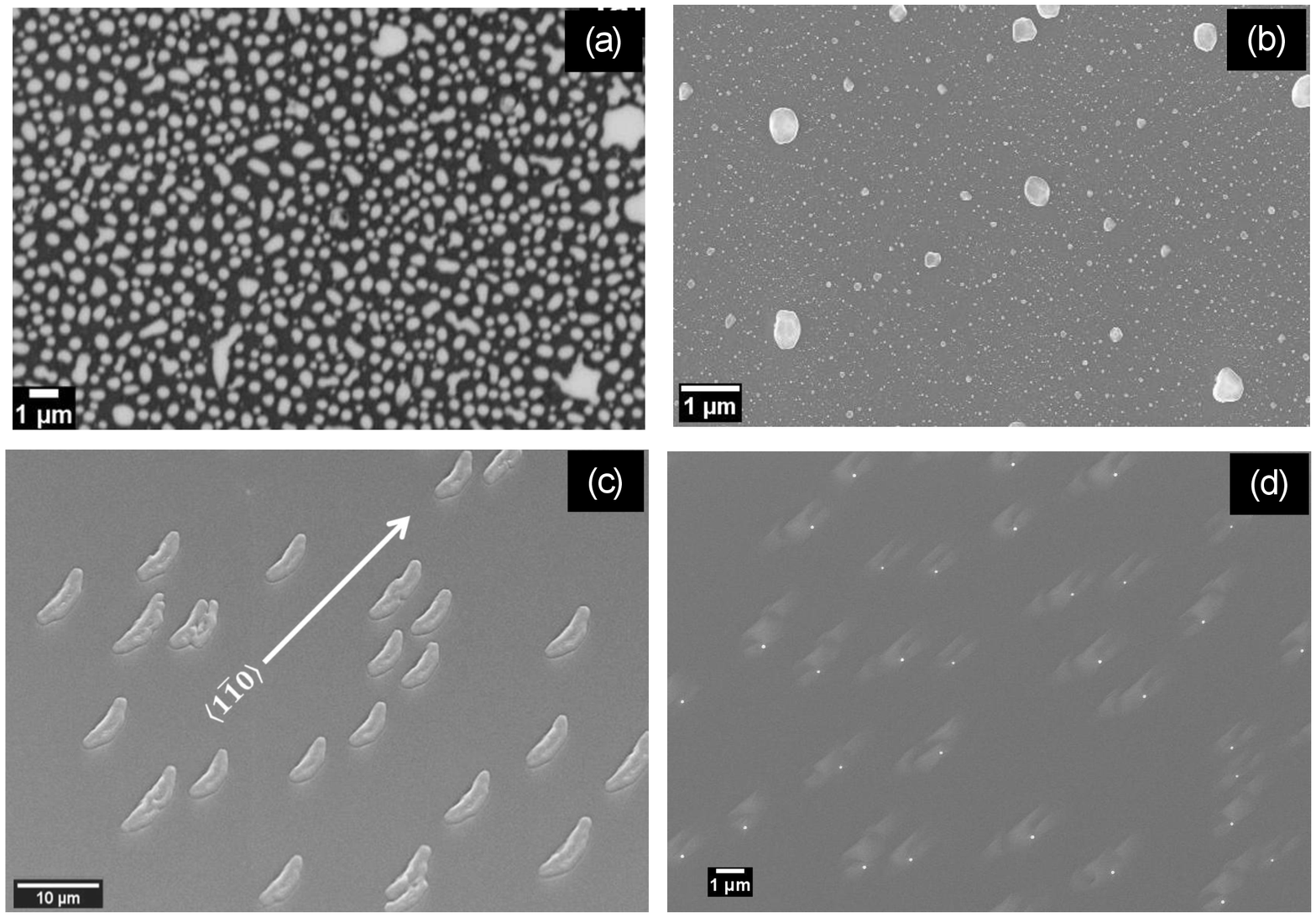, width=\linewidth}
\caption{ SEM micrographs  of 10 ML Ge/Si(5 5 12) at (a) RT
(b) 400$^{\circ}$C - DH (c) 600$^{\circ}$C - DH
(d) 800$^{\circ}$C - DH}\label{fig5}
\end{figure}

\begin{figure}
\centering  \vspace{0.4cm} \epsfig{file=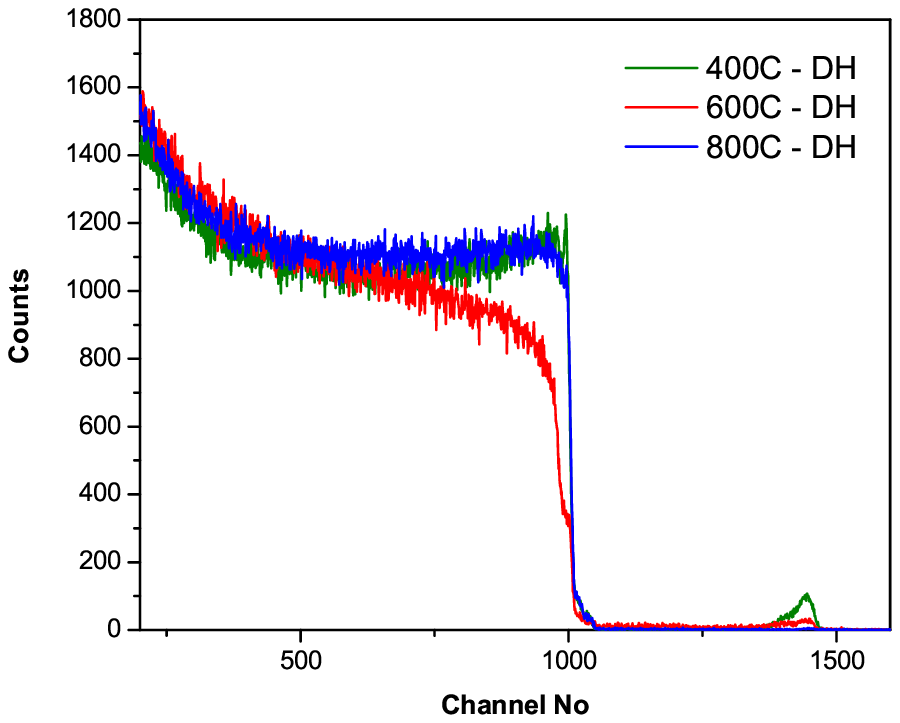, width=\linewidth}
\caption{Experimental RBS spectrum from 10 ML Ge/Si (5512) at T =
400$^\circ$C, 600$^\circ$C and 800$^\circ$C in DH
case.}\label{fig6}
\end{figure}

\begin{figure}
\centering  \vspace{0.4cm} \epsfig{file=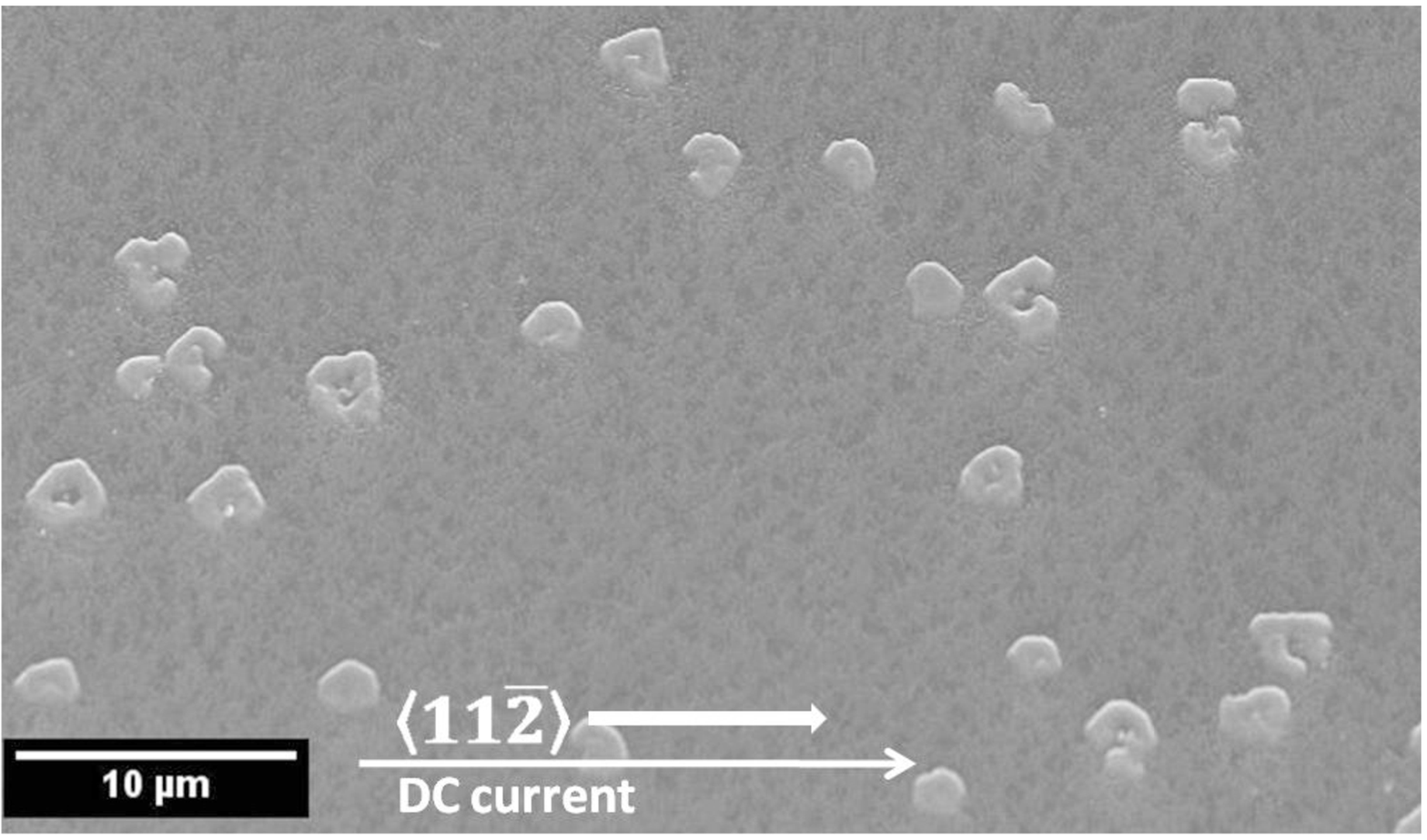, width=\linewidth}
\caption{SEM micrographs of germanium islands formed during the
10 ML Ge deposition on Si(111) for the DC heating applied along
$\langle 1 1 \bar{2}\rangle$. Irregular  structurs are observed with no proper alignment.}\label{fig7}
\end{figure}

We have discussed the temperature dependent shape evolution of
Ge-Si islands in the radiative heating (RH) condition. In this
section, we investigate the morphological change with increasing
temperature in the DC heated condition. Here, the DC current direction was parallel to the step direction i.e. $\langle 1\bar{1}0\rangle$. Fig. \ref{fig5} is a
SEM micrograph of Ge-Si structures at various temperatures in
the DH case. When substrate temperature T = 400$^\circ$C [where the applied current was 1.1 A and voltage 5.8 V] , the
 Ge-Si islands starts nucleating to form random  bigger islands with the reduction of smaller island density.
 The mean size of the bigger islands is 645.0 $\pm$ 22.0 nm. On further increasing
temperature to 800$^\circ$C [applied 1.8 A and 5.2 V] and above, the Ge-Si islands start
evaporating and hence the island density decreases.
 The signal of Ge diffusion in this case can be seen from the RBS spectrum in fig.\ref{fig6}.
The interesting shape transition occurs at T= 600$^\circ$C by passing the direct current through the sample with 1.5 A and 5.3 V . Here, well
aligned Si$_{1-x}$Ge$_x$ trapezoidal microstrures of average
length 6.25 $\pm$ 0.27 $\mu$m and aspect ratio 3.13 $\pm$ 0.3 form
along the step edge $\langle 1 \bar{1} 0 \rangle$ [fig. \ref{fig5}(c)]. In fig. \ref{fig6}, the RBS measurement has  revealed
the intermixing of Ge and Si. Additionally, the inward diffusion
of Ge (channels 1250-1350) into bulk Si and the outward diffusion
of Si (channels 900-940) can be clearly seen from
fig. \ref{fig6}. The RBS simulation has been done treating the
specimen as a multilayer with variation of Si and Ge
concentrations in each layer [table \ref{tab:table2}]. We found a
graded composition of Ge-Si in these structures. The
synchrotron-based high resolution x-ray diffraction (HRXRD) also
showed presence of graded Si$_{1-x}$Ge$_x$ composition from the
interface to the top of the trapezoid structures(data not shown).\\

\begin{table}
\caption{\label{tab:table2} Parameters obtained on fitting the
experimental RBS data for Ge/Si(5 5 12) at T=600$^\circ$C- DH,
using the SIMNRA simulation code.}
\begin{indented}
\item[]\begin{tabular}{@{}llll}
\br
 Layer &Ge  &Si  &Thickness \\
number&composition&composition&(1$\times$10$^{15}$
atoms.cm$^{-2}$)\\
\mr
 1 & 0.7500 & 0.2500 & 0.75\\
 2& 0.0750 & 0.9250 & 2.0\\
 3& 0.0070 & 0.9930 & 150.0\\
 4&0.0050 & 0.9950 & 250.0\\
 5&0.0040 & 0.9960 & 350.0\\
 6& 0.0020 &0.9980 & 500.0\\
 7& 0.0010 & 0.9990 & 3500.0\\
 8& 0 & 1 & Bulk\\
 \br
\end{tabular}
\end{indented}
\end{table}

In the DH condition, DC current acted as an additional
factor in stimulating the motion of germanium adatoms towards the
edges of steps. In the ref.\cite{ronspies}, R$\ddot{o}$nspies
\emph{et al} showed that steps in a high index surface, apparently
do not interrupt the conducting path, but due to local scattering
at step edge  increase the local resistivity by more than one
order of magnitude. Hence, the local temperature along the step
direction is more and stimulate the Ge-Si diffusion along the
steps making the structures well elongated.

To see the effect of direct current on low index silicon surfaces, we performed similar growth and characterization for Ge/Si(111) system. Fig. \ref{fig7} depicts the SEM micrograph for 10 ML deposited Ge film on Si(111) surface in DH condition. In this case, the DC direction was applied along $\langle 1 1 \bar{2}\rangle$. Here irregular structures are formed with no proper alignment.

\section{Conclusion}
In summary, we report the shape evolution of MBE grown Si$_{1-
x}$Ge$_x$ islands on reconstructed high index Si(5 5 12)  surfaces, as a function of growth temperature.
Also, the mode of heating (i.e. RH and DH)  plays a vital role in
the shape transformations. We show that a self assembled growth at
optimum temperature leads to interesting shape transformations,
namely, spherical islands to faceted dome structures in case of RH
condition and  to elongated trapezoidal structures for the DH
case. We have also observed  the intermixing of Si-Ge in the
larger faceted dome structures and aligned trapezoidal
structures.We look forward to carry out a spectroscopic analysis
along the step edges and of the aligned Si$_{1- x}$Ge$_x$ structures as well.\\

\ack
 PVS would like to thank the Department of Atomic Energy, Government of India for
granting FEGSEM under 11th plan.

\end{document}